# Differentiation of Sliding Rescaled Ranges: New Approach to Encrypted and VPN Traffic Detection

Raoul Nigmatullin
Radioelectronics and Informative
Measurement Technics Department
Kazan National Research Technical
University named after A.N. Tupolev,
(KNRTU-KAI)
Kazan, Russia
renigmat@gmail.com

Alexander Ivchenko
Laboratory of Multimedia Systems and
Technologies
Moscow Institute of Physics and
Technology
Dolgoprudny, Russia
ivchenko.a.v@phystech.edu

Semyon Dorokhin
Laboratory of Multimedia Systems and
Technologies
Moscow Institute of Physics and
Technology
Dolgoprudny, Russia
dorohin.sv@phystech.edu

*Abstract* — **We propose a new approach to traffic preprocessing called Differentiation of Sliding Rescaled Ranges (DSRR) expanding the ideas laid down by H.E. Hurst. We apply proposed approach on the characterizing encrypted and unencrypted traffic on the well-known ISCXVPN2016 dataset. We deploy DSRR for flow-base features and then solve the task VPN vs nonVPN with basic machine learning models. With DSRR and Random Forest, we obtain 0.971 Precision, 0.969 Recall and improve this result to 0.976 using statistical analysis of features in comparison with Neural Network approach that gives 0.93 Precision via 2D-CNN. The proposed method and the results can be found at https://github.com/AleksandrIvchenko/dsrr_vpn_nonvpn.**

## I. INTRODUCTION

Encryption on the Internet is ubiquitous being the foundation for secure communications. Creation, development and protection of applications make traffic classification a challenge for the research community. The traffic classification of individual applications is necessary for its further processing: network management; network QoS (Quality of Service) and user QoE (Quality of Experience) assessment and compliance; security; accounting.

To define and process distinct application traffic, it is necessary to separate it from the general flow. In the case of unencrypted traffic, it is possible to define an application by parsing headers. However, it induces latency at each node and reduces quality in terms of QoS and QoE. Furthermore, significant computational resources are required to analyze large traffic flows. However, such an approach is not possible for encrypted traffic.

Due to the variety of protocols, applications and qualitatively different traffic of stationary and mobile devices, an effective method [15] is to select encrypted traffic first (traffic characterization) and then classify the traffic within this subset.

In this paper, we focus on analyzing regular encrypted traffic and encrypted traffic tunneled through a Virtual Private Network (VPN). VPN-tunnels are used to secure data confidentiality of a network connection. Since VPN encryption is applied at network level, it is extremely difficult to identify applications functioning via VPN. That is why VPN is commonly used to obtain access to information prohibited by local authorities.

We considered the time-related traffic features as numerical series and developed a new method: Differentiation of Sliding Rescaled Ranges (DSRR). We used DSRR for preprocessing and afterwards applied basic ML (machine learning) algorithms. It leads significant gain in terms of ML and surpassed DL (deep learning) without heavy computing operations.

In Section II, we consider the approaches and results of the analysis of other researchers on the considered dataset. Section III is devoted to the description of the proposed data preprocessing technique. Section IV presents the results of our approach in context of basic machine learning methods and provides a tabular comparison with other works. In Section V, the application of statistical analysis is presented, which allowed improving the quality of the model. At the end of the article, plans for further research are given.

## II. STATE OF THE ART

The initial traffic classification techniques associated transport layer ports with specific applications. Nevertheless, this approach has low accuracy and low reliability, thus Deep Packet Inspection (DPI) appeared. The DPI approach analyzes packets and solves the classification problem using signatures or patterns. However, DPI techniques that require payload examination are not computationally efficient, especially over a high-bandwidth network. Moreover, their quality drops sharply in the presence of encapsulation and encryption.

Choosing effective and reliable features or parameters for traffic analysis is still a major challenge. The classification of network traffic is mainly divided into two types [6]: flow-based classification, using properties such as flow bytes per second, duration per flow, etc. and packet-based classification, using properties such as size, inter-packet duration of the first or several packets, etc.

In this paper, we focus on the well-known flow-based dataset created by Gil et al. [6] at 2016. This dataset consists of time-related features to avoid dependence of encryption type. We investigate the scenario represents the VPN vs Non-VPN traffic.

Since the release, the data under consideration have been studied in more than a hundred works. These works show the effectiveness of ML (machine learning) and DL (deep learning) for traffic classification in general due to a large number of types of protocols and applications and features, which, moreover, are constantly updated. The proliferation of mobile devices has also brought in a huge variety of traffic.

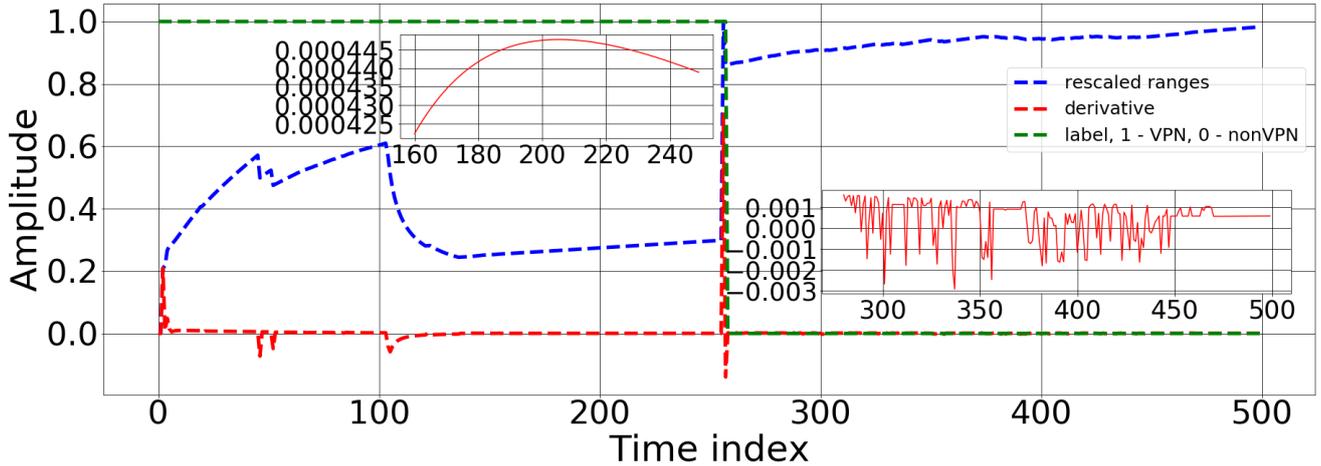

Fig. 1. Demonstration the difference between VPN (left) and NonVPN (right) preprocessed feature "duration" applying the SSR approach

In the original paper [6], the basic ML methods C 4.5 and kNN are investigated on the entire set of extracted features, and common metrics Precision (Pr) or Positive Predictive value and Recall (Rc) or Sensitivity are used to assess the quality. The best results are achieved using the C4.5 algorithm: Pr 0.89 for VPN and 0.906 for NonVPN [6].

Wang et al. [7] applied DL-approach. Neural network 2D-CNN shows the result according to the Pr metric of 89%. Aceto et al. [8] explored the application of various DL methods and considered them in detail: Stacked Auto Encoder, LSTM, 1D-CNN, 2D-CNN Hybrid LSTM + 2D-CNN.

At the papers [8, 9, 10] input data employed to feed the DL traffic classifier is characterized by the protocol layer ("ALL" vs. "L7") and the TC object ("Flow" vs. "Biflow") considered, with "Biflow +ALL" input combination achieving the best performance. Unfortunately, such design choice led to biased results (for a comparison taking into account this issue see [11]).

The dataset under investigation [6] is used to test Deep Packet [9] and Datanet [10], two DL-based encrypted traffic classifiers working at packet-level and adopting a 1D/2D-CNN, a (deep) MLP or a Stacked AutoEncoder. The Deep Packet [9] uses metrics Precision, Recall and F1-measure. In VPN-nonVPN classification problem, Deep Packet demonstrates Precision 0.93 with CNN. Datanet [10], as well as in [9], uses preprocessing and balancing the dataset, the same metrics, but only for the classification of individual protocols within the class of encrypted traffic.

There is a well-known approach to time series analysis based on rescaled range, which is described in [2]. Range $R(n)$ and standard deviation $S(n)$ are calculated in a window of size $n$. Rescaled range $R(n)/S(n)$ further calculated for different values of $n$. For long-range dependent processes, rescaled range has an exponential dependency on $n$, i.e.

$$\frac{R(n)}{S(n)} \approx Cn^H \qquad (1)$$

$H$ is called Hurst exponent and can be used as a characteristic of a process' degree of long-range dependency. Though Hurst exponent was originally used to describe storage capacity of reservoirs, it is commonly used to analyze time series of various nature, etc. for financial time series [4].

It is tempting to analyze whether internet traffic obeys the same law. Hurst exponent was used for DDoS attack detection with satisfactory results [5], but the described method requires quite a large number of packets to provide reliable results. In this paper, we propose a novel approach based on rescaled range.

III. THE PROPOSED METHOD

The new approach can be interpreted as a preprocessing method for machine learning algorithms. Let us suppose there is a time series of labels $y(t_i)$, $i=0,…,N-1$ and $F$ time series of features corresponding to it: $x_j(t_i)$, $j=0,…,F-1$. Features have clear interpretation (table 1), for instance, $x_j(t_i)$ may be the mean value of active flow time. However, according to preliminary calculations on DDoS dataset [13], this approach is also suitable for packet-based features.

The proposed algorithm transforms features $x_j(t_i)$ to features $\tilde{x}_j(t_i)$ using the following procedure. The entire time scale $t_i$ is divided into blocks of size $w < N$. For every block $w$ the following algorithm is applied (Fig. 2):

1. Calculate rescaled range for first *a* samples within the block
2. Calculate rescaled range for first *2a* samples within the block

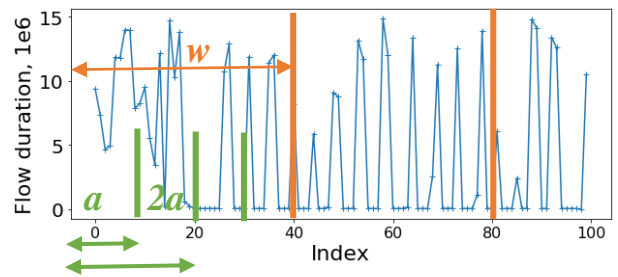

Fig. 2. Hurst algorithm intuition, $w = 40$, $a = 10$

3. The same for *3a* samples.
4. Continue till rescaled range is calculated over the entire block *w*

For simplicity, we will use $a = 1$. In that case for every block of length *w* one obtains rescaled ranges curve of length *w*. Rescaled ranges curve is further differentiated using a simple first-order computational scheme and the derivative is used as a new feature.

Thus, the original features $x_j(t_i)$ are mapped to new ones $x_j(t_i)$. The idea of this approach is based on the assumption that encrypted and unencrypted traffic is qualitatively and quantitatively different; therefore, there are bursts or breaks at their junctions. The derivative used as a simple measure if such discontinuities, as it is shown in Fig. 1.

For non-VPN traffic the derivative oscillates near zeros and has no outliers. In case of VPN traffic, there two possible scenarios: (a) the derivative either has a significant peak which corresponds to a fast change of the original feature, (b) or it remains small compared to non-VPN case because there was a major discontinuity before and therefore *R(n)* remains constant with only *S(n)* varying in (1). It seems that for VPN traffic the derivative is either extremely large (fast change of a feature) or extremely low (range is constant due to prior fast feature change, only variance changes).

The analysis of the SRR curves of the approach on 23 pairs of VPN features does not give an inside that one or several pairs are significantly different in every time interval. Nevertheless, the combination of all SRR-features gives a significant improvement of prediction.

In this case, we suppose that the reason is connected with local explanation of the combination of group of features. In the next paper, we plan to analyze the importances of features using the LIME approach [14]. In the next sections, we will demonstrate that this approach allows to increase in terms of Accuracy, Pr and Rc in the task of traffic characterization.

## IV. DSRR AND MACHEINE LEARNING

The dataset considers 8 main features, which are used to calculate 23 indicators, using maximums, minimums, averages and standard deviations for individual connections (Table I). Data flow is defined by a sequence of packets with the same values for {Source IP, Destination IP, Source Port, Destination Port and Protocol (TCP or UDP)}. The duration of flows are 15, 30, 60 and 120 seconds.

TABLE I. FEATURES DESCRIPTION

| Feature | Table Column Head |
|---|---|
| duration | Duration of a flow |
| fiat | Forward Inter Arrival Time, time between two packets sent forward (mean, min, max, std) |
| biat | Backward Inter Arrival Time, time between two packets sent backward (mean, min, max, std) |
| flowiat | Flow Inter Arrival Time, time between two packets sent in either direction (mean, min, max, std) |
| active | Amount of time flow was active before going idle (mean, min, max, std) |
| idle | Amount of time flow was idle before going active (mean, min, max, std) |
| fb_psec | Flow bytes per second |
| fp_psec | Flow packets per second |

As Wang et al. [1] did, we also noticed a slight decrease in the quality of the model with an increase in the time interval, getting the best results at 60 seconds' flow using Random Forest with 0.971 Pr (Table II). We divide the dataset on stratified train and test parts with the ratio 7:3. We use well-known sklearn library and Python 3.7.

TABLE II. THE COMPARISON BETWEEN OUR METHOD, OTHER RESLULST AND BASIC ML

| Method (Algorithm) | Pr | Rc | F1 |
|---|---|---|---|
| Random Forest + DSRR + Correlation Analysis | **0.976** | **0.973** | **0.974** |
| Random Forest + DSRR | **0.971** | **0.969** | **0.970** |
| Random Forest | 0.80 | 0.80 | 0.79 |
| Gil et al. [6], C4.5 | 0.89 | * | * |
| Deep Packet [9], CNN | 0.93 | * | * |
| Wang et al. [7], CNN | 0.89 | * | * |
| Decision tree + DSRR | 0.93 | 0.93 | 0.93 |
| Decision tree | 0.75 | 0.75 | 0.74 |
| kNN + DSRR | 0.91 | 0.91 | 0.91 |
| kNN | 0.81 | 0.80 | 0.80 |

* means "not mentioned".

The Fig. 3 illustrates how much gain in terms of accuracy we have with the proposed preprocessing method: there is an improvement from 0.895 to 0.967 on the 60s subset.

## V. CORRELATION ANALYSIS

Using universal correlation coefficient $\Phi_\kappa$ [12] with normalized value from 0 to 1, we investigate the level of association of features and the binary target (VPN or nonVPN), and evaluate the monotonic relationship between features. None of the parameters in isolation have a pronounced association with the target.

The maximum association with label VPN/nonVPN according to $\Phi_\kappa$ are following: max_biat – 0.18, total fiat – 0.16 total_biat – 0.15 min_fiat – 0.1. Others no more than 0.07. However, there are associations close to the maximum between the features. We removed features with $\Phi_\kappa = 1$. In terms of Kendall Tau correlation coefficient we deleted features with correlation more than 0.87 (Fig.4).

Thus, we have the combination of Random Forest, DSRR and Correlation Analysis that gives 0.976 Pr.

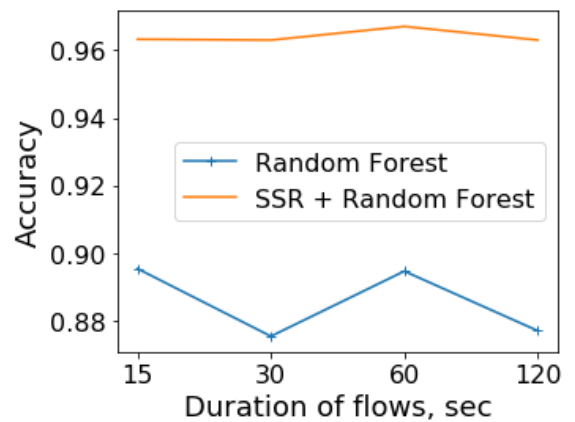

Fig.3. Accuracy scores gain with applying DSRR on the dataset

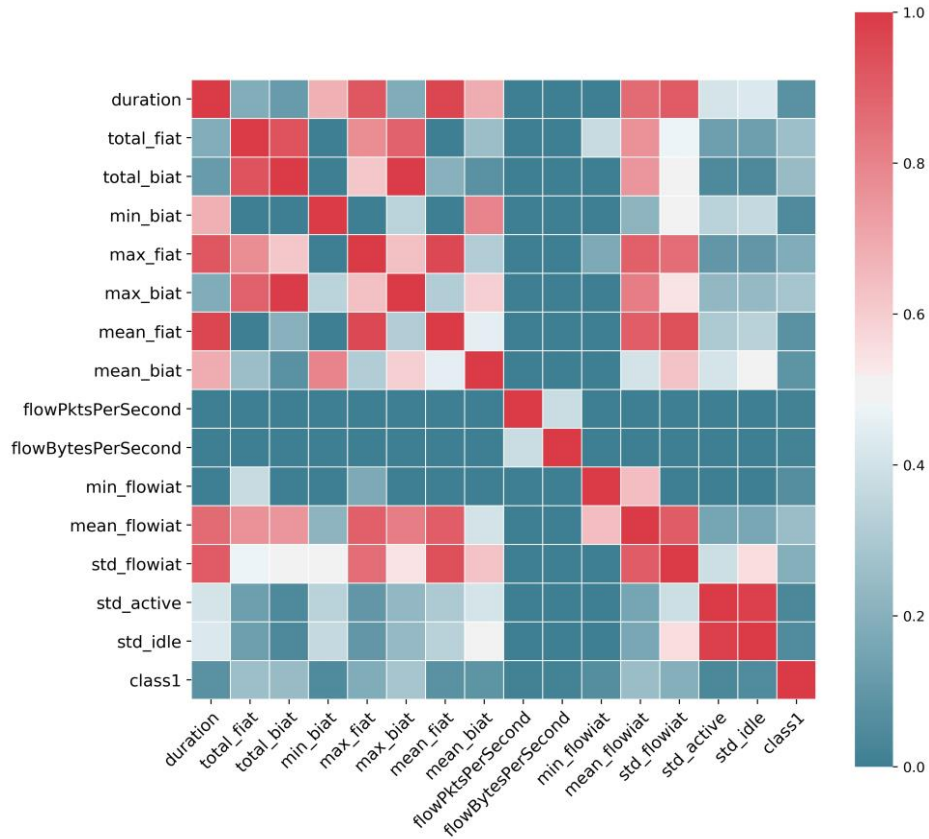

Fig. 4. Heat map using *Фк* correlation coefficient. Red color means high level of association. We cleared the features from pairwise correlations and improved the Random Forest model prediction.

## VI. RESULTS AND DISCUSSION

We have applied the DSRR approach to transform the time-based features of network traffic. We used the assumption that there are bursts in the features, which the derivative should detect efficiently. In addition, a statistical analysis of the features using *Фк* correlation coefficient was carried out and strong mutual pairwise correlations were removed.

This approach allowed us to obtain the result on the Random Forest model up to 0.976 Precision and 0.973 Recall. Our computationally simple mathematics outperform heavy neural network 2D-CNN in the last papers [7, 9] and shows the best result on this dataset so far.

The main drawback of the proposed method is that it can transform feature in a block-wise fashion only, which will induce a detection delay dependent on the length of the window.

In further research, we will explore DSRR approach, using various combinations of block sizes *w* and sample counts *a*, including block overlays. Furthermore, we will solve the traffic classification task to identify individual services and protocols within encrypted traffic.

This paper did not investigate the issues of data processing speed and comparison of processing time with other models, however, given the heavy neural network calculations and data preprocessing, we can assume the proposed method has significantly smaller computational complexity.